\documentclass[%
 aip,
 amsmath,amssymb,
 reprint,%
fleqn]{revtex4-2}

\usepackage{graphicx}
\usepackage{dcolumn}
\usepackage{bm}
\usepackage{color}

\usepackage[utf8]{inputenc}
\usepackage[T1]{fontenc}
\usepackage{mathptmx}
\usepackage{amsmath,amssymb}
\usepackage{etoolbox}
\usepackage{mathtools}
\usepackage{scalerel}
\usepackage{amsmath} 
\usepackage{amssymb}
\usepackage{amsfonts} 
\usepackage{graphicx} 
\usepackage{epstopdf}
\usepackage{gensymb}
\usepackage{enumerate}
\usepackage{color}
\usepackage{amsthm}
\usepackage{tcolorbox}
\usepackage{braket}
\usepackage{xcolor}
\usepackage{float}
\usepackage{multirow}
\tcbuselibrary{theorems}
\usepackage[thinc]{esdiff}

\newtcbtheorem[]{mytheo}{ }%
{colback=gray!5,colframe=gray!35!black,fonttitle=\bfseries}{th}
\DeclareMathOperator\erf{erf}

\usepackage{makeidx}
\usepackage[utf8]{inputenc}
\usepackage[english]{babel}
\usepackage{hyperref}
\hypersetup{
    colorlinks=true,
    linkcolor=blue,
    filecolor=magenta, 
    citecolor=blue,
     urlcolor=cyan,}

\newcommand{\beq}{\begin{equation}}
\newcommand{\eeq}{\end{equation}}

\makeatletter
\def\@email#1#2{%
 \endgroup
 \patchcmd{\titleblock@produce}
  {\frontmatter@RRAPformat}
  {\frontmatter@RRAPformat{\produce@RRAP{*#1\href{mailto:#2}{#2}}}\frontmatter@RRAPformat}
  {}{}
}%
\makeatother

\bibpunct{}{}{,}{s}{}{\textsuperscript{,}}
\begin{document}


\title{Computational projects with the Landau-Zener problem in the quantum mechanics classroom}

\author{Livia A. J. Guttieres}
\affiliation{Department of Physics,
             University of Chicago, 5720 South Ellis Avenue,
             Chicago, IL 60637 USA}
\author{Marko D. Petrovi\'c}
\author{James K. Freericks}%
 \email{james.freericks@georgetown.edu}
\affiliation{Department of Physics, Georgetown University,
             37th and O Sts. NW, Washington, DC 20057 USA}

\date{\today}

\begin{abstract}
The Landau-Zener
problem, where a minimum energy separation is passed with
constant rate in a two-state quantum-mechanical system, is
an excellent model quantum system for a computational project. It
requires a low-level computational effort, but has a number of
complex numerical and algorithmic issues that can be resolved
through dedicated work.
It can be used to teach
computational concepts such as accuracy, discretization, and
extrapolation, and it reinforces quantum concepts of time-evolution
via a time-ordered product and of extrapolation to infinite time
via time-dependent perturbation theory.  In addition, we discuss the concept
of compression algorithms, which are employed in many advanced
quantum computing strategies, and easy
to illustrate with the Landau-Zener problem.
\end{abstract}

\maketitle
\section{\label{sec:level1}Introduction}


The Landau-Zener problem is an exactly solvable problem in
quantum mechanics that describes how a quantum particle tunnels
between two states as a function of the speed with which it
traverses an avoided crossing.\cite{landau,zener}~ The exact
solution involves mapping the time-dependent Schr\"odinger
equation onto the so-called Weber equation, which is solved with
parabolic cylinder functions. But because these functions are
not so familiar to most students, this mapping is rarely taught.
Instead, because the system is just a two-state system, one can
compute the results numerically. This brings in issues related
to discretization and to accuracy, which can be particularly
acute for high accuracy because the solution has slowly
decaying oscillations that make determining the final tunneling
probability challenging without invoking some form of averaging.
Instead, one can use time-dependent perturbation theory to
append the long-time results and achieve much higher
accuracy solutions. 

This makes the Landau-Zener problem an excellent choice for a
computational project in a quantum mechanics
class. The time evolution, via a Trotter product formula, is
easy to code. Appending the time evolution at long times requires a
mastery of time-dependent perturbation theory and the
interaction representation. Modifying the discretization size
and the time cutoff for the time evolution allows students to
understand issues related to the accuracy of the computation.
Finally, this specific problem has a few different compression
strategies that can be employed---these strategies replace the
product of a string of operators by a single operator exactly
equal to the product. Compression strategies are employed in
quantum computing to reduce the depth of a quantum
circuit.\cite{compression1,compression2}~ Here, one can learn
 how such compression strategies work and how to parameterize SU(2) rotations in two
different ways to complete the compression. 

In this work, we describe a student-led
project on the Landau-Zener problem that will enable students to
learn many of these different topics related to quantum
mechanics and computation. This can be achieved even with
beginner to intermediate competency with programming because
the codes required are quite simple to implement. It also
provides a nice mix between formal development and computational
work, similar to  much of contemporary research.

The Landau-Zener problem was originally solved in 1932 by Landau,\cite{landau} Zener,\cite{zener} Stueckelberg,\cite{stuckelberg} and Majorana.\cite{majorana}~ 
We also have found it discussed in two textbooks: Konishi and Pafutto~\cite{konishi-pafutto} and Zweibach.~\cite{zweibach} Interestingly, the Landau-Zener problem is not widely discussed in other quantum mechanics textbooks, even though
it is ubiquitous in modern physics. Historically, it was initially applied to inelastic atomic and molecular collisions. Beyond collisions, two-level systems that exhibit  nonadiabatic transitions include Rydberg atoms in rapidly rising electric fields, qubit states in an NV center in diamond, and double-quantum dots. Other systems include qubits based on Josephson junctions, charge qubits in semiconductor quantum dots, graphene devices with an avoided crossing near the Dirac point, ultracold molecules in a laser trap, and even time-resolved photoemission in charge-density-wave systems. A discussion of many of these applications is given in a recent review article.\cite{review}

The problem has also been discussed in the pedagogical
literature. One study explores the accuracy of Runge-Kutta integration of the Schr\"odinger equation,\cite{Ya-Feng-Cao}~ while another uses the Landau-Zener approximation,\cite{LZA}~ 
which turns out not to be very accurate. The problem is mapped to the problem of a sphere rolling without slipping\cite{sphere}~ and solved classically, and it is also solved using a simple conceptual approximation that averages probabilities, not probability amplitudes, due to the fast oscillations.\cite{simple}~ Finally, another approach uses contour integrals.\cite{formula} Our work focuses on developing a computational project that employs perturbation theory, compression, and numerical evaluation of the time-ordered product to explore the interesting physics and numerics.

The remainder of the paper is organized as follows. In Sec.~II,
we introduce the Landau-Zener problem and discuss the Pauli spin
matrix identities needed to work with the problem. In Sec.~III,
we describe the discretized time evolution via the Trotter product formula.
In Sec.~IV, we illustrate how time-dependent perturbation theory
can append the time evolution of the
semi-infinite tails using the interaction picture in a
first-order expansion. Compression algorithms are discussed in
Sec.~{V}, followed by implementation strategies for the classroom  and conclusions in Sec.~{VI}.
\section{The Landau-Zener problem and Pauli spin-matrix
         identities}



\begin{figure}[tb]
\includegraphics[width=0.49\textwidth]{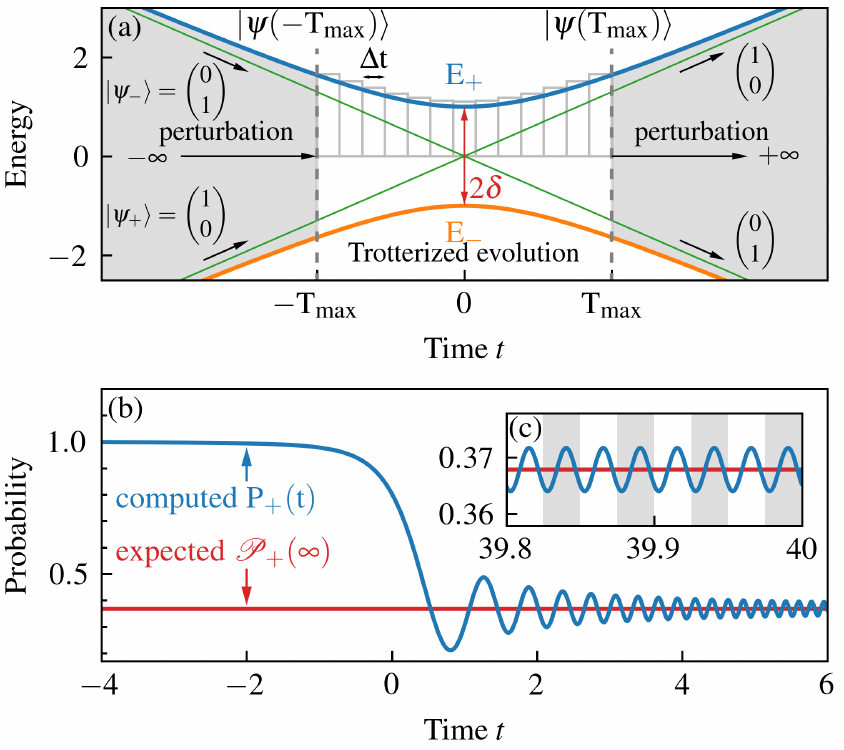}
\caption{\label{fig:LZ}
  (Color online) Landau-Zener system: (a) Time evolution of the two
  instantaneous eigenenergy levels $E_{\pm}(t) = \pm
  \sqrt{v^2t^2 + \delta^2}$ (the upper blue and the lower orange
  curve). At time $t = 0$, the two levels are separated by the
  minimal energy gap of $E_+(0) - E_{-}(0) = 2\delta$. The green
  lines show the crossing of the ($\delta=0$) energy levels $\pm vt$ 
  at time $t = 0$ in a system that has no coupling between the levels.  The time evolution algorithm that we use is divided
  into three parts. The initial state $|\psi(-T_{\rm
  max})\rangle$ is obtained either using perturbation theory
  applied to $|\psi_{+}\rangle$ at $t=-\infty$
  (what we consider as \textit{the perturbed state}), or is set to
  $|\psi_{+}\rangle$ at $-T_{\rm max}$ (what we consider as \textit{the
  unperturbed initial state}). This initial state is then propagated
  towards $|\psi(T_{\rm max})\rangle$ using an evolution
  operator in the Trotterized form with a time step of $\Delta
  t$. An additional perturbation is applied to $|\psi(T_{\rm
  max})\rangle$ to obtain the final state at $t = +\infty$, or
  $|\psi(T_{\rm max})\rangle$ is considered to be the final
  state (for the two different types of calculations).
  (b) Time evolution of the computed transition probability
  $P_{+}(t) = |\langle\psi_{+}|\psi(t)\rangle|^2$ compared
  to the expected probability $\mathcal{P}_{+}(\infty)$ 
  obtained from
  the analytical expression for the Landau-Zener transition. The rate
  is $v = \pi$ and $\delta = 1$; we use these same parameters for all of the numerical calculations in this paper. The inset in
  panel (c) shows the fast oscillations of the transition
  probability $P_{+}(t)$ with time on a backdrop of width
  $vt^2 = 4N\pi$ (gray and white background), showing that these
  oscillations have a period proportional to $\sim t^2$. The
  amplitude of these oscillations for the interval $[-T_{\rm
  max},T_{\rm max}]$ decays quite slowly with $T_{\rm max}$ (as a power law)
  unless corrected by the time-dependent perturbation theory.
 }
\end{figure}

The Landau-Zener problem consists of determining the 
probability to transition
from the ground-state to the excited state of a two-level system, after the two states
approach each other with an avoided crossing and then depart from each other.
The Landau-Zener system is described by the Hamiltonian 
\begin{align}
    \label{eqn:land}
    \hat{H}(t) = \begin{pmatrix}
                   v t & \delta \\
                   \delta & -v t 
            \end{pmatrix} = vt\sigma_z + \delta\sigma_x,
\end{align}
where $t$ is time, $v$ is the rate at which the two levels
approach each other, and $\delta$ is the coupling between them
that determines the minimal energy gap ($2\delta$) of the avoided crossing (occurring at time $t = 0$). 
Both $v$ and $\delta$ are real numbers with units of energy/time and energy, respectively. The symbols $\sigma_z$ and
$\sigma_x$ represent Pauli spin matrices
\begin{equation}
  \sigma_x = \begin{pmatrix}
              0 & 1 \\
              1 &0 
             \end{pmatrix},\ \
  \sigma_y = \begin{pmatrix}
              0 & -i \\
              i & 0
             \end{pmatrix},\ \
  \sigma_z = \begin{pmatrix}
             1 & 0 \\
             0 & -1
             \end{pmatrix}.
\end{equation}
By diagonalizing the Landau-Zener Hamiltonian 
using time as a \textit{parameter},
one obtains two instantaneous eigenenergy levels shown in
Fig.~\ref{fig:LZ}(a). Initially, when $t =
-\infty$, the two energy eigenvectors $|\psi_\pm\rangle$ given by
\begin{equation}
    |\psi_+\rangle=\begin{pmatrix}1\\0\end{pmatrix}~~\text{and}~~|\psi_-\rangle=\begin{pmatrix}0\\1\end{pmatrix},
\end{equation}
are infinitely 
separated in energy and the system starts in
the ground-state
$|\psi(-\infty)\rangle = |\psi_{+}\rangle$. 
As the system evolves with
time, the two levels $E_{+}(t)$ (the upper instantaneous energy level) and $E_{-}(t)$ (the lower instantaneous energy level) approach each
other as $t\to 0$ and then move apart as $t\to\infty$. Because the ground state smoothly changes from $|\psi_+\rangle$ as $t\to-\infty$ to $|\psi_-\rangle$ as $t\to\infty$,  the probability to remain in the ground state for large positive times is given by
$P_{-}(t) = |\langle \psi_{-}|\psi(t)\rangle|^2$. Similarly, the probability to end in the
excited state at long times is given by
$P_{+}(t) = 1 - P_{-}(t) = |\langle \psi_{+}|\psi(t)\rangle|^2$.
We are
interested in both of these probabilities when $t\to\infty$, $P_{+}(\infty)$ and $P_{-}(\infty)$.

As shown in Fig.~\ref{fig:LZ}(a), to transition from the lower-energy state to the higher-energy state, when traversing
the avoided crossing, the system has to tunnel through
a gap of size at least $2\delta$, so the probability
to transition is associated with a tunneling process.
Depending on the value of the rate $v$ and the level
separation $\delta$, we distinguish between two types of
transitions. If the system evolves adiabatically, that is,
extremely slowly,
it will always remain in the lower-energy state (the lower band);
that is, it will make perfect transitions from one instantaneous ground state to another along its time evolution.
According to the diagram in Fig.~\ref{fig:LZ}(a) this means that
around $t=0$ there will be a slow and smooth transition from 
$|\psi_+\rangle$ to $|\psi_-\rangle$. If we let the system evolve
diabatically (fast), it tunnels from the lower to the upper band. The objective of 
the Landau-Zener calculation
is to precisely determine
these probabilities as functions of $v$ and $\delta$.

Zener\cite{zener} solved 
the full time-dependent 
problem analytically by mapping
the equation of motion of the system into the form of the Weber
equation, which allowed him to obtain the exact solution  $\mathcal{P}_{+}(\infty) =
\exp\left(-\pi\delta^2/v\right)$. Here, 
we present a computational approach to find the same solution
numerically. 

The first issue that arises when considering this problem
from a numerical perspective is how to deal with the infinite times. Numerical simulations work with finite times, 
so how does one effectively start with a
state at $t =-\infty$ and obtain a result at $t = +\infty$
on a computer? It is usually assumed that starting
in the state $|\psi_{+}\rangle$ at some sufficiently large
(but finite) negative time is justified, and will lead to an accurate
numerical solution.  But, the Landau-Zener problem is known for its slowly
decaying oscillations of the transition probability $P_{+}(t)$
with time, which are illustrated on the right hand side of
Fig.~\ref{fig:LZ}(b) and in Fig.~\ref{fig:LZ}(c).
Although the time evolution of a two-level system is
a simple problem to solve numerically and is not 
computationally demanding, the persistence of these slowly
decaying oscillations presents a serious problem in accurately determining the transition probability as $t\to\infty$.
This problem might seem trivial here, since the analytical
solution is known, 
but it becomes important for generalizations of the Landau-Zener problem, where the level separation is not linear
in time and the exact solution is not known. In our numerical approach, we show how the
time evolution can be divided into three 
parts, where propagation from $-\infty$ to some cutoff time and from another cutoff time to
$+\infty$ can be resolved using time-dependent perturbation
theory (see the gray areas in Fig.~\ref{fig:LZ}(a)), whereas the
evolution on the finite time interval between the cutoffs can be computed 
using the Trotter product formula, which discretizes the time
evolution operator. We discuss both these approaches in more
detail in the following sections.

Before we explain how to implement these quantum-mechanical
concepts, we have to establish some mathematical prerequisites
necessary to understand time evolution in quantum mechanics and
time-ordered products of Hamiltonians based on Pauli matrices (two-level systems).
Pauli matrices satisfy the commutation relations
\begin{equation}
 [\sigma_i,\sigma_j]= \sigma_i\sigma_j - \sigma_j\sigma_i =
  2i\sum_k\epsilon_{ijk}\sigma_k,
\end{equation}
where the
indices $i$, $j$ and $k$ represent the 
coordinates $x$, $y$, and $z$, the factor $2i$ is twice the imaginary number $i$,  and
$\epsilon_{ijk}$ is the Levi-Civita (completely antisymmetric)
tensor;
this is a tensor that is equal to 1 when $ijk$ is an even permutation of 123, is equal to $-1$ when $ijk$ is an odd permutation of 123, and vanishes otherwise.
Similarly, their anticommutator is given by
\begin{equation}
  \{\sigma_i,\sigma_j\}=\sigma_i\sigma_j + \sigma_j\sigma_i =
   2\delta_{ij}\mathbb{I},
\end{equation}
where $\mathbb{I}$ is the unit matrix and $\delta_{ij}$ is
the Kronecker delta function.
These two expressions can be combined to create the product
formula for any two Pauli matrices
\begin{equation}
    \sigma_i\sigma_j=\delta_{ij}
    \mathbb{I}+i\sum_k\epsilon_{ijk}\sigma_k.
\end{equation}
The product formula is useful for evaluating the
exponentials of weighted sums of Pauli matrices, which are needed to construct the time evolution
operators. The exponential of a linear combination of Pauli
matrices can be expanded into an infinite series
\begin{align}
  e^{i\vec{\gamma}\cdot\vec{\sigma}} = 
    \sum_{n=0}^{\infty} \frac{(i\vec{\gamma}
    \cdot\vec{\sigma})^n}{n!}.
\end{align}
Here $\vec{\gamma}$
is a 3-component vector of real numbers and the dot product is
understood as 
$\vec{\gamma}\cdot\vec{\sigma} = \gamma_x\sigma_x +
\gamma_y\sigma_y + \gamma_z\sigma_z$ (which is a $2\times 2$ matrix).
The quadratic term $(i\vec{\gamma}\cdot\vec{\sigma})^2$
in the series is computed by using the product formula for two Pauli matrices. One obtains
\begin{align}
    (i\vec{\gamma}\cdot\vec{\sigma})^2
    =
    -\sum_i \sum_j\gamma_i \gamma_j\sigma_i \sigma_j
    =
    -\sum_i \gamma_i \gamma_i \mathbb{I}=-|\gamma|^2\mathbb{I}.
\end{align} 
Here, the $\sum_{ij}\epsilon_{ijk}\gamma_i\gamma_j\sigma_k$ term 
is zero because 
one can interchange the $i$ and $j$ indices in the summation and show that
$\epsilon_{ijk}\gamma_{i}\gamma_{j}\sigma_{k} = 
-\epsilon_{ijk}\gamma_{i}\gamma_{j}\sigma_{k}$.
The infinite sum can then be broken into two sums---those involving even powers and those involving odd
powers. Each can be resummed to yield $\cos{|\gamma|}$ or 
$\sin{|\gamma|}$.
This simplification yields the generalized Euler identity for 
Pauli matrices
\begin{equation}
\label{eqn:exp}
  e^{i\vec{\gamma} \cdot \vec{\sigma}} = 
  \cos|\vec{\gamma}|\,\mathbb{I}+i\sin|\vec{\gamma}|
  \frac{\vec{\gamma}\cdot\vec{\sigma}
  }{|\vec{\gamma}|},
\end{equation}
which transforms the symbolic expression for an exponential of
the Pauli matrices into a concrete $2\times 2$ matrix. 
This result will be employed in computing the time-evolution operator. We also use the product formula of two Pauli matrices to compute the product of two exponentials of linear combinations of Pauli matrices via
\begin{align}
  \label{eq:prod}
  e^{i\vec{\gamma}\cdot\vec{\sigma}}
  e^{i\vec{\gamma}\,'\cdot\vec{\sigma}}
  & = \left(
      \cos|\vec{\gamma}|\cos|\vec{\gamma}\,'|
      -\sin|\vec{\gamma}|\sin|\vec{\gamma}\,'|
      \frac{\vec{\gamma}\cdot\vec{\gamma}\,'}{%
       |\vec{\gamma}||\vec{\gamma}\,'|}
      \right)
      \mathbb{I}\nonumber\\
  & +i\left(
      \cos|\vec{\gamma}|\sin|\vec{\gamma}\,'|
      \frac{\vec{\gamma}\,'}{|\vec{\gamma}\,'|}
      +
      \sin|\vec{\gamma}|\cos|\vec{\gamma}\,'|
      \frac{\vec{\gamma}}{|\vec{\gamma}|}
      \ \right. \nonumber \\
  &
    \left.
     \phantom{+i\left(.\right)}
      -\sin|\vec{\gamma}|\sin|\vec{\gamma}\,'|
      \frac{\vec{\gamma}\times \vec{\gamma}\,'}{%
            |\vec{\gamma}||\vec{\gamma}\,'|}
    \right)\cdot\vec{\sigma}.
\end{align}
It is important to emphasize that, unlike exponentials of
real numbers, the expression 
$e^{i\vec{\gamma}\cdot\vec{\sigma}}
e^{i\vec{\gamma}\,'\cdot\vec{\sigma}}
=
e^{i\left(\vec{\gamma}+\vec{\gamma}\,'\right)\cdot\vec{\sigma}}
$ does not generally apply for exponentials of Pauli matrices.
The reason for this is the last term in Eq.~(\ref{eq:prod}) with
the scalar triple product. When this term is present,
the product of the two exponentials does not commute
and they can not be interchanged. In the case when
two vectors $\vec{\gamma}$ and $\vec{\gamma}\,'$ are colinear, then the triple-product term vanishes, the
two coefficients can be summed, and the two exponentials do
commute.

\section{Computational approaches to the time-ordered product}


Regardless of how we propagate from $t=-\infty$ to $t=-T_{\rm max}$ (and $t=T_{\rm max}$ to $t=\infty$), we still must use the computer to explicitly propagate from $t=-T_{\rm max}$ to $t=T_{\rm max}$. 
To time evolve the  state $|\psi(-T_{\rm max})\rangle$
to the state $|\psi(T_{\rm max})\rangle$, we must
apply the appropriate time-evolution operator. The time-evolution operator satisfies
\begin{align}
    \ket{\psi (t)}=\hat{U}(t,t_0)\ket{\psi (t_0)},
\end{align}
and depends on the initial time $t_0$ and the final time
$t$. The time-evolution operator can be found from the facts
that it must be unitary (so that it preserves the norm of the
state at any time $|\langle\psi(t)|\psi(t)\rangle|^2 =
1$) and that it must have the ``semi-group property,''
which implies that time evolution is additive. In other words, evolving the state from $t_0\to t^{\prime}$ and from
$t^{\prime}\to t$ is the same as directly
evolving the state from $t_0\to t$. Since the Hamiltonian must govern the time evolution, we are led to $\hat{U}(t,t_0)=e^{-\tfrac{i}{\hbar}(t-t_0)\hat{H}}$ for time-independent Hamiltonians because the sign in the exponent is determined by convention. 

In the time-dependent case, we determine the full evolution operator by considering the time evolution over a short time interval from $t$ to $t+\Delta t$. We simply assume that the time interval is short enough that we can take $\hat{H}(t)$ as being piecewise constant (over the time interval of length $\Delta t$) and use the constant Hamiltonian time evolution operator for this
piecewise constant Hamiltonian over the short interval $\Delta t$. Then 
\begin{align}
\label{eqn:ham}
  \hat{U}(t+\Delta t, t) = 
  e^{-\tfrac{i}{\hbar} 
  \hat{H}\left(t\right)\Delta t}.
\end{align}
By applying a sequence of these operators
one can construct a discretized version of the evolution 
operator
\begin{align}
  \hat{U}(t,t_0) 
  = &\ \hat{U}(t,t-\Delta t)\hat{U}(t-\Delta
    t,t-2\Delta t)\cdots \nonumber \\
    & \cdots \hat{U}(t_0+2\Delta
    t,t_0+\Delta t)\hat{U}(t_0+\Delta t,t_0),
\end{align}
which approaches the exact evolution operator as 
$\Delta t\rightarrow 0$. 
This is called the Trotter product formula. In the limit
of $\Delta t \rightarrow 0$, the time-ordered product is conventionally written as 
\begin{equation}
  \label{eq:order}
  \hat{U}(t, t_0) = \mathcal{T}
  \left[\exp(-\frac{i}{\hbar}\int_{t_0}^{t}\hat{H}(t')dt')
  \right]
\end{equation}
where $\mathcal{T}$ is the time-ordering operator, which orders times with the ``latest times to the left.''

In our numerical 
approach to the Landau-Zener problem, the time evolution  over the finite interval $-T_{\rm max}\le t\le T_{\rm max}$ is performed via the
Trotter product formula with the time step  
$\Delta t$. 
We write the evolution operator
$\hat{U}(T_{\rm max}, -T_{\rm max})$ as a time-ordered 
sequence of exponentials of the Landau-Zener Hamiltonian
\begin{align}
  \label{eq:whole}
  \hat{U}(t+\Delta t, t) &= 
  e^{-\frac{i}{\hbar}\left(vt\sigma_z +
  \delta \sigma_x\right)\Delta t}=\cos\left (\sqrt{v^2t^2+\delta^2}\,\tfrac{\Delta t}{\hbar}\right )\mathbb{I}\nonumber\\
  &-i\frac{\sin\left (\sqrt{v^2t^2+\delta^2}\,\tfrac{\Delta t}{\hbar}\right )}{\sqrt{v^2t^2+\delta^2}\,\tfrac{\Delta t}{\hbar}}\left (vt\sigma_z+\delta\sigma_x\right )\tfrac{\Delta t}{\hbar},
\end{align}
after using the generalized Euler identity for the 
exponentials of the Pauli matrices in Eq.~(\ref{eqn:exp}). Each of 
the exponential Trotter factors $U(t+\Delta t,t)$ is now a concrete $2\times 2$ matrix, that acts on the state at a given time. Here, we use the exponentiated linear superposition of Pauli matrices with the vector 
$\vec{\gamma} = -\frac{\Delta t}{\hbar}
                  \left(vt, 0, \delta\right)^{T}$
(which is a time-dependent vector) for each Trotter factor.
Note that in general 
$\exp\left[i(a\sigma_z + b\sigma_x)\right]
\neq \exp(ia\sigma_z)\exp(ib\sigma_x)$ 
because the matrices
$\sigma_z$ and $\sigma_x$ do not commute, as explained in
the previous section. However, if the coefficients $a$ and $b$
are very small, as in the case of Eq.~(\ref{eq:whole}) for
$\Delta t\rightarrow 0$, then
$\exp\left[i(a\sigma_z + b\sigma_x)\right]
\approx \exp(ia\sigma_z)\exp(ib\sigma_x)$ because the error
corresponding to the commutator of the two terms is on the order
of $\sim \Delta t^2$. Thus if the evolution operator
is written in the Trotter product form, when $\Delta t$ is
sufficiently small, one can approximate a Trotter factor via
\begin{equation}
  \label{eq:split}
  \hat{U}(t+\Delta t, t) \approx
  e^{-\frac{i}{\hbar}vt\sigma_z\Delta t}
  e^{-\frac{i}{\hbar}\delta \sigma_x\Delta t},
\end{equation}
which we call the {\it split}-form of the evolution operator. 
We contrast this to the evolution operator in 
Eq.~(\ref{eq:whole}), that we call the {\it exact} form since there
is no approximation (except assuming the Hamiltonian is constant over the time interval $\Delta t$).
The evolution operator expressed in the split form is accurate
to the order of $\Delta t^2$, so the two forms should approach
one another in the limit of $\Delta t \rightarrow 0$. 

The operator in the split form has an additional interesting
feature. The exponentials of the $\sigma_x$ matrix
$\exp(-\frac{i}{\hbar}\delta\sigma_x\Delta t)$ in the Trotter
product sequence are
constant for a fixed time step $\Delta t$,
whereas the $\sigma_z$ terms 
$\exp{\left(-\frac{i}{\hbar}vt\sigma_z\Delta t\right)}$
depend on time $t$. At special times, $T_{\rm n}$, 
the exponent will satisfy the condition 
$vT_{\rm n}\Delta t/\hbar = 2n\pi$. At these times, the
given Trotter component with $\sigma_z$ (in the split form) is a unit matrix. The next
$\sigma_z$ component in the Trotter product will have the
exponent $v(T_{\rm n}+\Delta t)\Delta t/\hbar$ which is the same
as the earlier term
$v(T_{\rm n-1} + \Delta t)\Delta t/\hbar$ and the same as the even earlier time
$v\Delta t\Delta t/\hbar$, and so on. This means that in the split form,
the Trotter product repeats with period  
$T = 2\pi\hbar/\left(v\Delta t\right)$. But we know that the exact time evolution operator is not periodic. So this periodicity is an artifact of using the split form. It means we must choose a $\Delta t$ such that the split form is not close to its periodic behavior over the interval $-T_{\rm max}\le t\le T_{\rm max}$. If we do not, then 
the accumulation of error in the split form leads to the transition probability 
repeatedly switching between $P_{+}(-T/2)$ and $P_{+}(T/2)$
as time passes through different $T_{\rm n}$ points. To obtain
precise results using the split method we should pick
$\Delta t$ so that $T > T_{\rm max}$, ideally 
$T_{\rm max}$ no larger than $T/4$.

We now discuss how to perform the numerical calculation. The time evolution
can be implemented in two different ways. One can write a
function that computes the time-evolution operator $\hat{U}(
t+\Delta t, t)$ at every time step and applies it to the state
$|\psi(t)\rangle$ to obtain $|\psi(t + \Delta t)\rangle$
(propagating the state), or one can multiply this time-evolution operator with the accumulated time-evolution
operator for all previous times 
$\hat{U}(t+\Delta t, -T_{\rm max}) = \hat{U}(t+\Delta t, t)
\hat{U}(t, -T_{\rm max})$ (propagating the operator). In the latter case, the state is obtained from 
$|\psi(t+\Delta t)\rangle = 
 \hat{U}(t+\Delta t, -T_{\rm max})|\psi(-T_{\rm max})\rangle$.

Using the Trotter product formula combined with the generalized
Euler identity for each Trotter factor
enables the numerical solution of the Landau-Zener problem on
the finite time interval $\left[-T_{\rm max}, T_{\rm max}\right]$.
Since we are working with $2\times 2$ matrices that have an explicit form for each time step, this time evolution is relatively straightforward to program. 
However, we still need to determine the initial state $|\psi(-T_{\rm max})\rangle$. In the next section, we explain how to include the time
evolution operator over the two semi-infinite intervals by
using the interaction picture and the time-dependent
perturbation theory.
\section{Extrapolation to infinite time with time-dependent
perturbation theory} 


The roadblock to analytically determining the evolution operator for the
Landau-Zener problem in Eq.~(\ref{eqn:land}) is the linear
time dependence of the $\sigma_z$ term and the fact that the two
Pauli matrices ($\sigma_z$ and $\sigma_x$) do not commute.
This makes the time-ordered product in
Eq.~(\ref{eq:order}) virtually impossible to solve analytically. 

For large 
positive and negative
times, we must use time-dependent perturbation theory. However while in conventional quantum instruction, the unperturbed Hamiltonian is always chosen to be the time independent piece and the perturbation is time dependent, here, the unperturbed part of the Hamiltonian is the large piece (the $\sigma_z$ piece for large $|t|$), and the perturbation is the constant piece (the $\sigma_x$ piece). So, we
split the Landau-Zener Hamiltonian into two parts: the
main time-dependent Hamiltonian 
$\hat{H}_{\rm 0}(t) = vt\sigma_z$ and the time-independent
perturbation $\hat{V} = \delta\sigma_x$. The evolution 
operator is then constructed in the interaction picture via
\begin{align}
  \label{eq:PT}
  \hat{U}(t, t_0) & = \hat{U}_{0}(t, t_0)
  \hat{U}_I(t, t_0) \nonumber \\
            & = 
  \hat{U}_{0}(t, t_0)\mathcal{T}\left[e^{-\frac{i}{\hbar}
  \int_{t_0}^{t} \hat{V}_{\rm I}(t')dt'}\right],
\end{align}
where $\hat{U}_{\rm 0}(t, t_0)$ is the evolution operator for
the unperturbed Hamiltonian $\hat{H}_{\rm 0}(t)$, and
$\hat{U}_I(t, t_0)$ is the evolution operator for the perturbation.
In the interaction picture, we have
$\hat{V}_{\rm I}(t) = \hat{U}_{\rm 0}^\dagger(t, t_0)\,
\hat{V}\,\hat{U}_{\rm 0}(t, t_0)$. The time ordered
product $\mathcal{T}$ in Eq.~(\ref{eq:PT}) is to be understood in
the same sense as in Eq.~(\ref{eq:order}), meaning that we write it as an ordered sequence of exponentials of
each term in the integrand $\hat{V}_{\rm I}(t)$. Note that Eq.~(\ref{eq:PT}) is typically not the exponential of the integral of the operator 
$\hat{V}_{\rm I}(t)$; this only occurs if the integrand commutes with itself for different times. 
Note further that the notion of time-dependent perturbation theory here arises because as $t$ goes to $\pm\infty$, the unperturbed piece
$\hat{H}_{\rm 0}(t)$ is much larger than
$\hat{V}$. We also emphasize that
Eq.~(\ref{eq:PT}) is exact and does not contain any
approximation (the perturbative approximation arises from Taylor expanding the evolution the operator $\hat{U}_I(t, t_0)$ as we show
below).

The evolution operator for the unperturbed Hamiltonian 
$\hat{H}_{\rm 0}(t)$ can be computed analytically because it
commutes with itself for all times. Hence, we can just integrate the unperturbed piece to find
$\hat{U}_{0}(t, t_0) = e^{-\frac{i}{\hbar}
\frac{1}{2}v(t^2 - t_0^2)\sigma_z}$. We next use this exact result
to determine the perturbation in the
interaction picture (which is a rotation of the $\sigma_x$ matrix about the $z$-axis). It becomes
\begin{align}
  \label{eq:vi}
  \hat{V}_{\rm I}(t) & = 
  e^{\frac{i}{\hbar}\frac{1}{2}v(t^2 - t_0^2)\sigma_z}
  \delta\sigma_x
  e^{-\frac{i}{\hbar}\frac{1}{2}v(t^2 - t_0^2)\sigma_z}
  \nonumber \\
   & = \delta\left(\sigma_x \cos\frac{v(t^2 - t_0^2)}{\hbar} -
    \sigma_y \sin
    \frac{v(t^2 - t_0^2)}{\hbar}
   \right),
\end{align}
which can be found by using the generalized Euler identity to determine each exponential factor and then multiplying the three matrices together.
The strategy of perturbation theory is to approximate the time-ordered product for 
$\hat{U}_I(t, t_0)$ in the evolution operator 
by
\begin{equation}
  \hat{U}(t, t_0) \approx \hat{U}_{0}(t, t_{0})
  \left(\mathbb{I} - \frac{i}{\hbar}
  \int_{t_0}^{t} \hat{V}_{\rm I}(t')dt'
  \right), 
\end{equation}
which is accurate for $t$ close to $t_0$, or when $\hat{V}_I$ is ``small.''
\begin{figure}[tb]
\begin{center}
\includegraphics[width=0.49\textwidth]{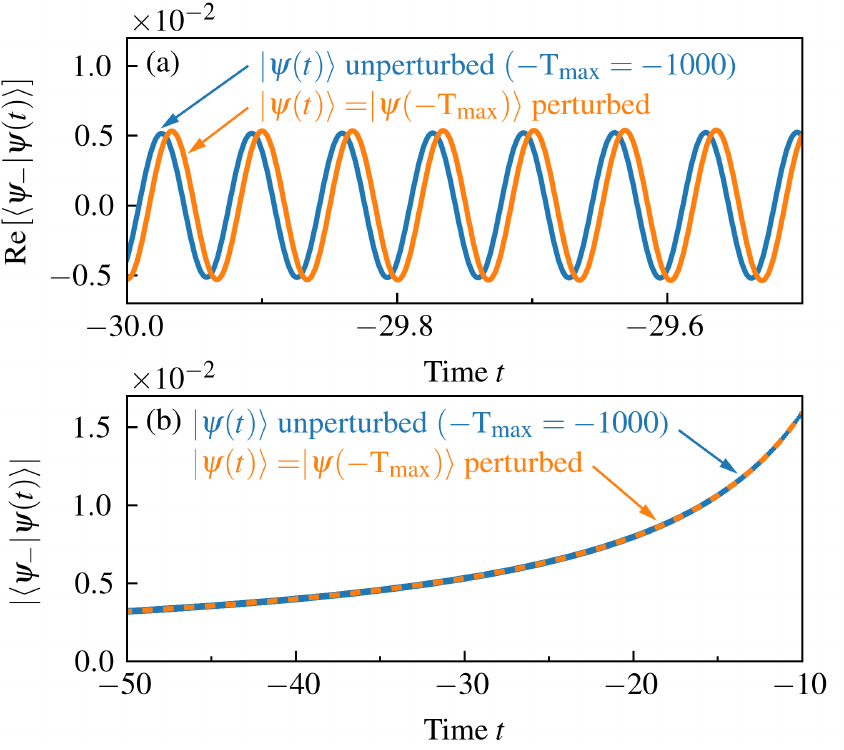}
\end{center}
\caption{\label{fig:PT}
 (Color online) State evolution for different starting conditions: (a)
 the real part of the projected wave function
 $\mathrm{Re}\left[\langle \psi_{-}|\psi(t)\rangle\right]$
 for the initial state computed as a function of time using 
  perturbation theory 
 $|\psi(t)\rangle = |\psi(-T_{\rm max})\rangle$ 
 (orange curve) versus the unperturbed state $|\psi(t)\rangle$
 computed by the time evolution from $|\psi_{+}\rangle$ starting
 at $-T_{\rm max} = -1000$ (blue curve). 
 (b) Time evolution of the modulus of the state projection 
 $|\langle\psi_{-}|\psi(t)\rangle|$ computed for the same two 
 starting conditions as in panel (a).
 }
\end{figure}
Directly computing the $\hat{V}_{\rm I}(t)$ operator from 
Eq.~(\ref{eq:vi}) and then factoring the result in terms of Pauli matrices gives
\begin{align}
\hat{V}_{\rm I}(t') 
  & = \begin{pmatrix}
        0 & e^{\frac{i}{\hbar}v(t'^2 - t_0^2)} \\
        e^{-\frac{i}{\hbar}v(t'^2 - t_0^2)} & 0
        \end{pmatrix} \nonumber \\
 & = 
      e^{-\frac{i}{\hbar}\frac{1}{2}vt_0^2\sigma_z}
      \begin{pmatrix}
      0 & e^{\frac{i}{\hbar}vt'^2 } \\
      e^{-\frac{i}{\hbar}vt'^2} & 0
      \end{pmatrix} 
      e^{\frac{i}{\hbar}\frac{1}{2}vt_0^2\sigma_z}
     \nonumber \\ 
 & = 
      \hat{Z}(t_0)\, 
      \hat{V}'_{\rm I}(t')\,
     \hat{Z}^{-1}(t_0),
\end{align}
where $\hat{Z}(t) = e^{-\frac{i}{\hbar}\frac{1}{2}vt^2\sigma_z}$.
The unit matrix can also be written as $\mathbb{I} = 
\hat{Z}(t_0)\,\mathbb{I}\,\hat{Z}^{-1}(t_0)$
and $\hat{U}_0(t, t_0) = \hat{Z}(t)\hat{Z}^{-1}(t_0)$,
so the evolution operator becomes
\begin{align}
  \label{eq:zz}
  \hat{U}(t, t_0) 
  & \approx \hat{Z}(t)\hat{Z}^{-1}(t_0)
  \left[ \hat{Z}(t_0)\mathbb{I}\hat{Z}^{-1}(t_0) 
         \phantom{\int_{t_0}^t}\right. \nonumber \\
  & \phantom{\approx \hat{Z}(t)\hat{Z}^{-1}(t)\left[\right] }
      \left. -\frac{i}{\hbar}
      \hat{Z}(t_0)\int_{t_0}^{t}\hat{V}'_{\rm I}(t')dt'
  Z^{-1}(t_0)
  \right] \nonumber \\
  & = \hat{Z}(t)
  \left[ \mathbb{I} - \frac{i}{\hbar}
  \int_{t_0}^{t}\hat{V}'_{\rm I}(t')dt'
  \right] 
  Z^{-1}(t_0).
\end{align}
Eq.~(\ref{eq:zz}) yields an approximate formula for the
evolution operator in the interaction picture for any two 
times. We now apply it 
to compute $\hat{U}(-T_{\rm max}, -\infty)$
and $\hat{U}(+\infty, T_{\rm max})$. In the first case,
the operator $\hat{Z}^{-1}(t_0=-\infty)$ on the right side of
Eq.~(\ref{eq:zz}) can be neglected because we start from an
eigenstate of the $\sigma_z$ operator (the $|\psi_+\rangle$
state) and
acting with $\hat{Z}^{-1}$ will just produce a global  complex phase,
which does not affect the probabilities. In the second
case, when computing $\hat{U}(+\infty, T_{\rm max})$
the operator on the left side of Eq.~(\ref{eq:zz})
(i.e.~$\hat{Z}(t=\infty)$) can be
neglected because we are interested only in the probability
$P_{+}(\infty)$ and that term also just contributes a complex
phase, which cancels out. The matrix integral in 
Eq.~(\ref{eq:zz}) can be analytically computed in both cases.
For $\hat{U}(-T_{\rm max}, -\infty)$ it equals
\begin{align}
 \mathbb{I} - \frac{i}{\hbar}
  \int_{-\infty}^{-T_{\rm max}}\hat{V}'_{\rm I}(t')dt'=
   \begin{pmatrix}
     1 & -\frac{i}{\hbar}\xi\left(T_{\rm max}\right) \\
     -\frac{i}{\hbar} \eta\left(T_{\rm max}\right) & 1
   \end{pmatrix},
\end{align}
where
\begin{equation}
  \eta(t) = \sqrt{\frac{\pi}{2v}}
  \frac{(i-1)}{2}
  \left[1-\erf\left(\sqrt{\frac{v}{2}}
        (i+1)t\right)
  \right],
\end{equation}
and  
\begin{equation}
 \xi(t) = 
   \sqrt{\frac{\pi}{2v}} \frac{(i+1)}{2}
    \left[1+ \erf\left(
              \sqrt{\frac{v}{2}}(i-1)t
    \right)\right].
\end{equation}
Both off-diagonal components of the perturbation matrix, 
$\eta(t)$ and $\xi(t)$, are expressed in terms of the error function
\begin{equation}
 \erf(z) = \frac{2}{\sqrt{\pi}}\int_0^z e^{-t^2}dt,
\end{equation}
with a complex argument.
The evolution operator obtained through perturbation theory is approximate and not necessarily unitary. This means the
quantum state must be normalized ``by hand'' after applying the approximate evolution operator. In other words, we renormalize both at time $t = -T_{\rm max}$ and at $t = \infty$.
Note that the central integral in Eq.~(\ref{eq:zz}) does not
change for $\hat{U}(+\infty, T_{\rm max})$, which follows from
$\hat{V}'_{\rm I}(-t) = \hat{V}'_{\rm I}(t)$ and
from the change of variables $t'\rightarrow -t'$, but the
$\hat{Z}(t)$ and $\hat{Z}^{-1}(t_0)$ operators do change.

The improvement in accuracy for the calculation that includes the time-dependent perturbation theory arises from its use of a more accurate initial
state at $-T_{\rm max}$. This improvement is
shown in Fig.~\ref{fig:PT}.  In Fig.~\ref{fig:PT}(a),
we compare the phase of the state $|\psi(t)\rangle$ obtained
by propagating $|\psi_{+}\rangle$ from a distant time 
$-T_{\rm max} = - 1000$ (blue curve) with the one obtained by 
perturbation theory as functions of time ($t = T_{\rm max}$,
the orange curve). In general, the amplitude 
$|\langle \psi_{-}|\psi(t)\rangle|$ in Fig.~\ref{fig:PT}(b)
matches the perturbed and
unperturbed state, but as shown in Fig.~\ref{fig:PT}(a) 
they differ by a phase. This
initial phase difference introduces an error for later times
that propagates through the Trotter product.
The advantage of the perturbation theory is that it
reduces the computational demand by reducing the 
time range $\left[-T_{\rm max}, T_{\rm max}\right]$ needed for the numerical calculation. Instead
of propagating the state from a very distant past,
one can obtain a precise result for a much smaller $T_{\rm
max}$.
The perturbation that brings the state
to $t=+\infty$ is even more important, because it removes the
small but persistent oscillations of the transition probability,
as we show in Sec.~{VI}. Note that discussing time-dependent perturbation theory provides a number of benefits to the students: (i) it shows them how to extrapolate solutions from finite to infinite time; (ii) it illustrates how to employ time-dependent perturbation theory in a nonstandard fashion; and (iii) it shows how small errors in the initial conditions can propagate in a calculation and affect results at later times if they are not properly addressed.
\section{Compression algorithms}

\begin{figure}[tb]
\begin{center}
\includegraphics[width=0.49\textwidth]{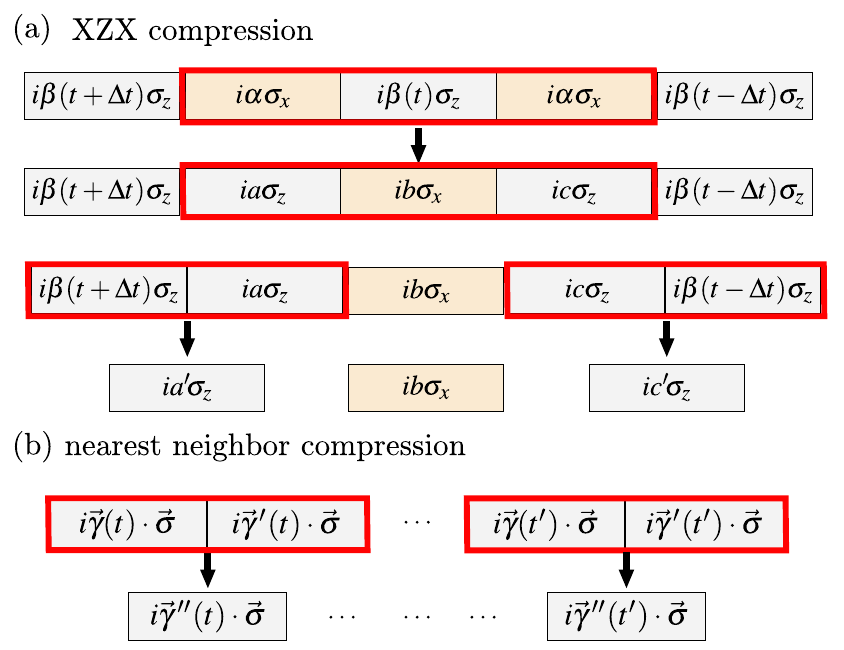}
\end{center}
\caption{\label{fig:XZX}
  Schematics of the two possible compression algorithms of the
  Trotterized evolution operator. The boxes represent
  exponential coefficients at every Trotter step. (a) $XZX$
  compression, where a $\sigma_x\sigma_z\sigma_x$ sequence
  of the evolution operator exponentials in the Trotter-split
  form is replaced with $\sigma_z\sigma_x\sigma_z$
  sequence, so that neighboring $\sigma_z$ components can be
  merged together. (b) The nearest neighbor compression method,
  where two neighboring Trotter components of the evolution
  operator (based on the exact Hamiltonian) are merged
  together at every compression step.}
\end{figure}

In Sec.~{III}, we explained how to compute the time evolution
operator with the Trotter product and time
discretization. Using the generalized Euler identity, the 
exponents at each time step can be converted into 
$2\times 2$
matrices and the evolution operator can be computed by 
matrix multiplication. Replacing a sequence of 
exponentiated operators (as in the Trotter product) 
with a single exponential operator is called {\it compression}. Sophisticated compression algorithms based on the Cartan decomposition of Lie groups are employed in quantum computing to greatly reduce the depth of circuits.\cite{compression1,compression2}~ 
In this section, we show how a simpler application using the two equivalent ways to represent rotations can be used to compress the time-evolution operator that we use in the Landau-Zener problem, which provides a nice opportunity to show how compression algorithms work in a quantum class.

Instead of computing the evolution operator by matrix
multiplication for each time step, compression relies
on combining the exponential parameters (the vectors
$\vec{\gamma}$ used in each exponential
$e^{i\vec{\gamma}\cdot\vec{\sigma}}$ factor) into a single vector,  for the evolution operator over the finite time interval. The basic idea for compression
comes from the fact that each exponent of a Pauli matrix
represents a rotation on the Bloch sphere. A sequence
of rotations about different axes can be replaced by a
single rotation around a single axis.  We focus on
two compression algorithms that could be used to
compute the evolution operator 
$\hat{U}(T_{\rm max}, -T_{\rm max})$.

The first algorithm is called $XZX$ compression and it is based
on a mathematical relationship between the Pauli matrices
\begin{equation}
  \label{eq:xzx}
  e^{-i\alpha\sigma_x} 
  e^{-i\beta\sigma_z} 
  e^{-i\gamma\sigma_x}
  =
  e^{-ia\sigma_z} 
  e^{-ib\sigma_x}
  e^{-ic\sigma_z},
\end{equation}
where 
one can compute $a$, $b$, and $c$ from the known coefficients $\alpha$, $\beta$, and $\gamma$. With the help of
the generalized Euler identity, we convert the left and the
right side of Eq.~(\ref{eq:xzx}) into $2\times 2$ matrices and
compute the relationships between the coefficients by equating
the four matrix elements of the final products on each side of the equation. This gives exact inverse trigonometric
relations
\begin{align}
  \label{eq:alpha}
  a = & \frac{1}{2} \arctan
      \left[\tan(\beta)\frac{\cos(\alpha-\gamma)}
                        {\cos(\alpha+\gamma)}\right]
      \nonumber\\
  &- \frac{1}{2} \arctan
  \left[\tan(\beta)\frac{\sin(\alpha-\gamma)}
                        {\sin(\alpha+\gamma)}\right],
\end{align}
\begin{align}
\label{eq:beta}
  b &= \arctan
  \left[\sqrt{\frac{\sin^2(\alpha+\gamma) +
  \tan^2(\beta)\sin^2(\alpha-\gamma)}{\cos^2(\alpha+\gamma) + 
  \tan^2(\beta)\cos^2(\alpha-\gamma)}}\right],
\end{align}
\begin{align}
  c = &
  \frac{1}{2} \arctan 
  \left[\tan(\beta)\frac{\cos(\alpha-\gamma)}
                        {\cos(\alpha+\gamma)}\right]\nonumber\\
  & + \frac{1}{2} \arctan
  \left[\tan(\beta)\frac{\sin(\alpha-\gamma)}
                        {\sin(\alpha+\gamma)}\right].
\end{align}
As shown in Fig.~\ref{fig:XZX}(a),
the identity in Eq.~(\ref{eq:xzx}) can be applied
to the time evolution operator for the Landau-Zener problem
written as a Trotter product in the split-form
in Eq.~(\ref{eq:split}).
Here, the Trotter product is a sequence of alternating
constant $\sigma_x$ and time-dependent $\sigma_z$ exponentials
(the first row of Fig.~\ref{fig:XZX}(a)). Taking
the three exponentials lying inside the red box in the first row of 
Fig.~\ref{fig:XZX}(a), we apply
Eq.~(\ref{eq:xzx}) and switch the ordering
of the operators to the one in the second row of
Fig.~\ref{fig:XZX}(a); that is, we change a product of exponentials in the form $XZX$ into the form $ZXZ$.
This re-expression of factors allows the now
adjacent $\sigma_z$ exponentials on the edges of the red box on the second line to merge into
a single exponential as emphasized by the red boxes in the
third row of Fig.~\ref{fig:XZX}(a). The initial five exponential
components are
then compressed into three. This procedure is repeated until
the Trotter product is reduced to just three exponential factors,
which can each be computed using the generalized Euler formula.

\begin{figure}[tb]
\begin{center}
\includegraphics[width=0.49\textwidth]{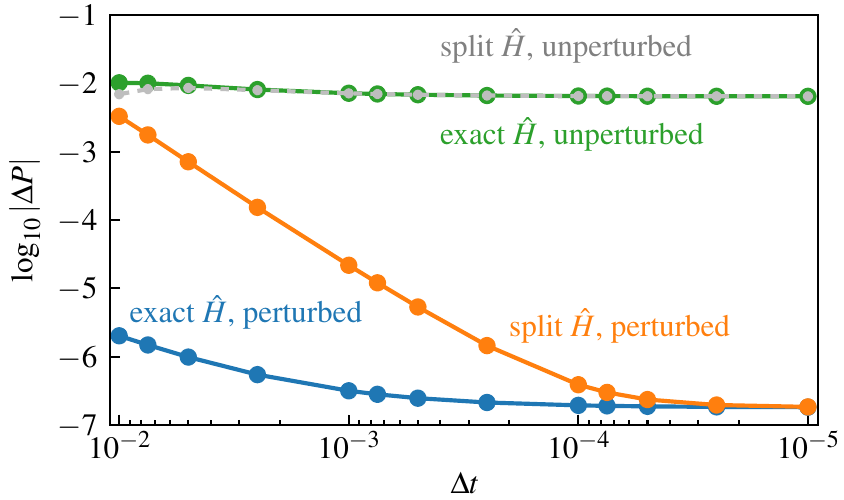}
\end{center}
\caption{\label{fig:DT}
 The accuracy of the computed transition probability as a function
 of the time step $\Delta t$. In the case of perturbed results,
 we compute the difference as
 $\Delta P = P_{+}(\infty) - \mathcal{P}_{+}(\infty)$,
 whereas in the case of unperturbed results we compute
 $\Delta P = P_{+}(T_{\rm max}) - \mathcal{P}_{+}(\infty)$.
 Here, $\mathcal{P}_{+}(\infty)$ is the probability obtained
 by the exact Landau-Zener formula $\exp(-\pi\delta^2/v)$, while $P_{+}(\infty)$ and
 $P_{+}(T_{\rm max})$ are the probabilities computed using
 the Trotterized evolution, with or without the perturbation,
 respectively. We also compare the accuracy when the  Trotter step exponential is evaluated exactly, or when the 
 Hamiltonian is split between $\sigma_z$, and $\sigma_x$ terms
 in each Trotter step. In all cases $T_{\rm max} = 30$. High accuracy can only be attained for this value of $T_{\rm max}$ when the infinite tails are included perturbatively.
 }
\end{figure}

The second compression algorithm is
computationally far more efficient than the XZX compression
since it does not require any inverse trigonometric functions.
We call this algorithm the nearest-neighbor algorithm because it
involves merging the neighboring
exponentials in the Trotter product as shown in
Fig.~\ref{fig:XZX}(b). In contrast to the XZX compression, which
requires the split form of the Trotter product, the nearest-neighbor compression uses an exact form for each Trotter factor.  This compression
algorithm is based on Eq.~(\ref{eq:prod}) for the product
of two exponentials
$e^{i\vec{\gamma}\cdot \vec{\sigma}}
e^{i\vec{\gamma}^{\ \prime}\cdot \vec{\sigma}}
= e^{i\vec{\gamma}^{\ \prime\prime}\cdot \vec{\sigma}}
$. We simply need to construct the
vector $\vec{\gamma}^{\ \prime\prime}$ from the known $\vec{\gamma}$ and
$\vec{\gamma}^{\ \prime}$. Comparing
the generalized Euler formula in Eq.~(\ref{eqn:exp}) and
its extension to a product of two exponentials in
Eq.~(\ref{eq:prod}) we immediately find that 
\begin{align}
  \label{eq:cos}
  & \cos|\vec{\gamma}\,''| = 
      \cos|\vec{\gamma}|\cos|\vec{\gamma}\,'|
      -\sin|\vec{\gamma}|\sin|\vec{\gamma}\,'|
      \frac{\vec{\gamma}\cdot\vec{\gamma}\,'}{%
       |\vec{\gamma}||\vec{\gamma}\,'|},
\end{align}
and
\begin{align}
  \label{eq:sin}
  \sin|\vec{\gamma}\,''|
  \frac{\vec{\gamma}\,''}{|\vec{\gamma}\,''|} = & \left(
      \cos|\vec{\gamma}|\sin|\vec{\gamma}\,'|
      \frac{\vec{\gamma}\,'}{|\vec{\gamma}\,'|}
      +
      \sin|\vec{\gamma}|\cos|\vec{\gamma}\,'|
      \frac{\vec{\gamma}}{|\vec{\gamma}|}
      \ \right. \nonumber \\
  & \left.
      \phantom{+i\left(.\right)}
      -\sin|\vec{\gamma}|\sin|\vec{\gamma}\,'|
      \frac{\vec{\gamma}\times \vec{\gamma}\,'}{%
            |\vec{\gamma}||\vec{\gamma}\,'|}
    \right).
\end{align}
\begin{figure}[tb]
\includegraphics[width=0.49\textwidth]{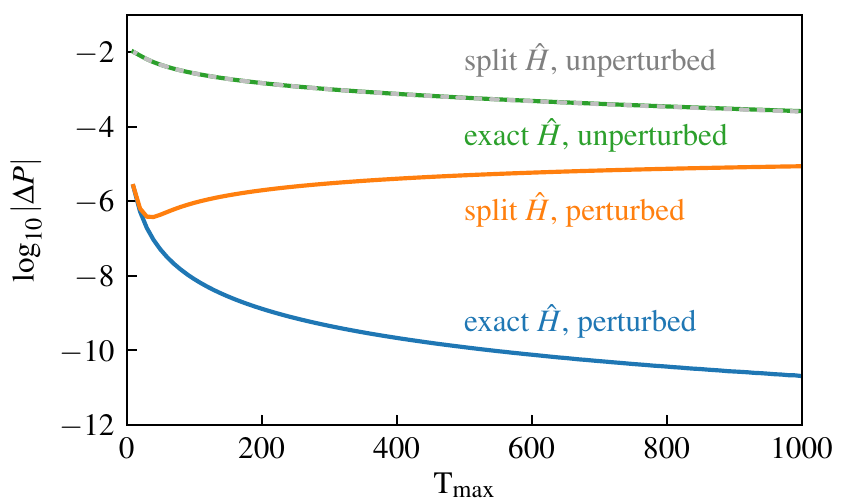}
\caption{\label{fig:TMAX}
  The accuracy of the computed transition probability, relative to the exact result $\exp(-\pi\delta^2/v)$, as a function
  of the finite time range $t\in[-T_{\rm max},
  T_{\rm max}]$. The time step  $\Delta t$ is fixed at the same value as
  in Fig.~\ref{fig:DT}, namely $\Delta t = 10^{-4}$. One expects that the time step $\Delta t$ must be reduced to accurately reach larger $T_{\rm max}$ values because of accumulated errors (here the $\Delta t$ value is held \textit{fixed}). This clearly occurs for the split method, which is much more sensitive to $\Delta t$ errors, but surprisingly is not impeding the exact, perturbed approach through nearly 12 digits of accuracy.
}
\end{figure}
The previous two equations connects both the four-component vector
$\left(\cos|\vec{\gamma}|,
\sin|\vec{\gamma}|\frac{\vec{\gamma}}{|\vec{\gamma}|}
\right)$
for $\vec{\gamma}$ and the one for $\vec{\gamma}\,'$ with the
one for $\vec{\gamma}\,''$. Nearest-neighbor compression
consists in computing these four-component vectors for every
step in the Trotter product and combining them using
Eq.~(\ref{eq:cos}) and Eq.~(\ref{eq:sin}). If the time interval
$\left[-T_{\rm max}, T_{\rm max}\right]$ is divided into
$2^{N_c}$ time steps, then the compression can reduce the
Trotter product to a single exponential in $N_c$ iterations (a logarithmic number of steps).
Computationally, this method is much faster than $XZX$ compression
and even the direct matrix multiplication, but it requires
a large amount of RAM memory for a small time step
$\Delta t$ because the four-component vectors are kept in memory
for every time step, whereas in the direct multiplication the
evolution operator is computed ``on the fly,`` which requires
storing only its current value. The memory requirement can be
reduced if the integration interval $\left[-T_{\rm max}, T_{\rm
max}\right]$ is divided into smaller intervals and compression
is applied to each one of them sequentially.

There are two benefits of this work for the students. First, they learn about the idea of compression, which is becoming increasingly important in quantum computing and second, they learn how one can revise initial algorithms to make them computationally more efficient, an important skill of the computational physicist.
\section{Implementation strategies and conclusion}

With these technical details completed, we now discuss how to implement this problem as a class project.
Figure~\ref{fig:DT} shows the numerical accuracy achieved using 
different approaches (perturbation vs.\ no perturbation and
exact vs.\ split
form of the evolution operator), expressed as an
error $\Delta P$ in determining the transition probability at
infinity on the logarithmic scale. The logarithmic scale tells
how many digits of accuracy one can achieve using different
approaches. The advantage of the time-dependent perturbation theory
is obvious since the unperturbed result converges
very slowly to the expected probability. The figure also shows
how the accuracy of the split-form of the Trotter product
approaches the accuracy of the exact Hamiltonian for 
sufficiently small $\Delta t$. The accuracy, in this case, is
limited by our choice of $T_{\rm max} = 30$ so the computed
results converge with decreasing $\Delta t$. Similarly,
Fig.~\ref{fig:TMAX}
shows how the integration range $[-T_{\rm max}, T_{\rm max}]$
influences the computed transition probability. The persistent
oscillations in the unperturbed system prevent determining the
transition probability beyond two or three digits of accuracy
even for $T_{\rm max} = 1000$. The slow decay of the unperturbed
error suggests that achieving higher accuracy in this case
is essentially impossible, even with increasing $T_{\rm max}$.
The perturbed results show a high accuracy
even for small $T_{\rm max}$. One way to increase the
accuracy in the unperturbed case is to do time-averaging once
$P_{+}(t)$ starts to oscillate around the expected
$\mathcal{P}_{+}(\infty)$ as done previously,\cite{Ya-Feng-Cao} but it is difficult to systematically do this when the amplitude is also decreasing with time, especially to high accuracy. We also can infer how the split approximation is more sensitive to $\Delta t$ errors, since we used a fixed time step, rather than reducing it as the cutoff time increases; here we see that the accumulated errors due to the finite size of the time step worsen the accuracy for large cutoff times. This advanced computational concept is beautifully illustrated in this project. Note that one need not worry about round-off error associated with matrix multiplications in this work. Those errors appear to always be much smaller than the other intrinsic errors of the computational algorithm.

The Landau-Zener problem is a challenging computational project for quantum-mechanics students, without requiring 
any knowledge of the higher level differential
equations that are usually used to solve this problem.
Students really get a taste of how computational physics works---they need to work through some nontrivial formalism to determine precisely what needs to be calculated and then they need to carefully program the results and run them. Finally, they need to examine the accuracy of the results. In the supplementary materials,  we provide a well-documented Python package that includes the codes 
used to produce the results presented in this paper. Teams can be formed to work on implementing
different approaches and collaborating to compare the different
outcomes (e.g perturbation vs.\ no perturbation or split Trotter
evolution vs.\ exact Trotter evolution) and discussing the
computational efficiency and the numerical accuracy of each of these approaches. 
The workload necessary to solve this problem goes beyond a
simple homework assignment, but it offers the opportunity for
students to gain deep knowledge of quantum mechanics in a practical
example that will allow them to acquire skills that are  important for
computational research.


\section{Acknowledgments}
L.A.J.G. was supported by the National Science Foundation
under Grant No.~DMR-1950502.
M.D.P. was supported by the Department of Energy,
Office of Science, Basic Energy Sciences
(BES) under Award DE-FG02-08ER46542.
J.K.F. was supported by the National Science Foundation
under Grant No. PHY-1915130 and by the
McDevitt bequest at Georgetown University. 
J.K.F. came up with the idea for the project, which was
examined in detail by L.A.J.G. and M.D.P. (M.D.P. acted as
the mentor to L.A.J.G.). All authors contributed to the
write up of the work.

\section{Author Declarations}
The authors have no conflicts to disclose.

\
\\
\\


\end{document}